\documentclass[aps,pre,reprint,groupedaddress]{revtex4-1}

\usepackage{amsmath, amssymb}
\usepackage{bm}
\usepackage{braket}
\usepackage{graphicx}

\usepackage[pdfencoding=auto]{hyperref}
\hypersetup{ %
  colorlinks=true, %
  linkcolor=blue, %
  citecolor=blue, %
  urlcolor=blue, %
}

\DeclareMathOperator{\sinc}{sinc}
\DeclareMathOperator{\Tr}{Tr}
\newcommand{\bTr}[2][]{\Tr_{#1}\! \left[ #2 \right]}
\newcommand{\Avr}[2][]{\left\langle #2 \right\rangle_{#1}}
\newcommand{\Abs}[1]{\left\lvert #1 \right\rvert}

\newcommand{\half}[1]{\frac{#1}{2}}
\newcommand{\inv}[1]{\frac{1}{#1}}

\newcommand{\leftscript}[2]{{\vphantom{#2}}#1\! #2}

\newcommand{\sKet}[2][]{\Ket{#2}_{#1}}
\newcommand{\sBraket}[4][]{\leftscript{_{#1}}{\Braket{#2 | #3 | #4}}_{#1}}
\newcommand{\sKetbra}[3][]{\Ket{#2}_{#1}\!\Bra{#3}}
\newcommand{\sProj}[2][]{\sKetbra[#1]{#2}{#2}}

\newcommand{\Commu}[2]{\left[ #1 , #2 \right]}
\newcommand{\dist}[3][]{d_{#1}\left( #2 , #3  \right)}
\newcommand{\Order}[1]{O\left( #1 \right)}

\newcommand{\xbk}[1]{{x\left( #1 \right)}}
\newcommand{\ybk}[1]{{y\left( #1 \right)}}

\newcommand{\calE}{\mathcal{E}}
\newcommand{\calK}{\mathcal{K}}

\newcommand{\Kmap}[2][]{{\calK#1}\!\left( #2 \right)}
\newcommand{\Emap}[2][]{{\calE#1}\!\left( #2 \right)}

\newcommand{\id}{\openone}
\newcommand{\Sys}{\text{S}}
\newcommand{\Ext}{\text{E}}
\newcommand{\Con}{\text{C}}
\newcommand{\Inter}{\text{I}}
\newcommand{\IE}{{\Inter\Ext}}
\newcommand{\SC}{{\Sys\Con}}

\newcommand{\hop}{\text{hop}}
\newcommand{\eff}{\text{eff}}
\newcommand{\tot}{\text{tot}}
\newcommand{\inter}{\text{int}}
\newcommand{\can}{\text{can}}
\newcommand{\init}{\text{i}}
\newcommand{\fin}{\text{f}}

\begin{document}

\title{Quantum Jarzynski equality of measurement-based work extraction}


\author{Yohei Morikuni}
\affiliation{Department of Physics, The University of Tokyo, Komaba, Meguro, Tokyo 153-8505}

\author{Hiroyasu Tajima}
\affiliation{Department of Physics, The University of Tokyo, Komaba, Meguro, Tokyo 153-8505}
\affiliation{Center for Emergent Matter Science (CEMS), RIKEN, Wako, Saitama 351-0198}

\author{Naomichi Hatano}
\affiliation{Institute of Industrial Science, The University of Tokyo, Komaba, Meguro, Tokyo 153-8505}


\date{\today}
\begin{abstract}
  Many studies of quantum-size heat engines assume that the dynamics of an internal system is unitary and that the extracted work is equal to the energy loss of the internal system.
  Both assumptions, however, should be under scrutiny.
  In the present paper, we analyze quantum-scale heat engines, employing the measurement-based formulation of the work extraction recently introduced by Hayashi and Tajima~[\href{http://arxiv.org/abs/1504.06150}{M. Hayashi and H. Tajima, arXiv:1504.06150}].
  We first demonstrate the inappropriateness of the unitary time evolution of the internal system (namely the first assumption above) using a simple two-level system; we show that the variance of the energy transferred to an external system diverges when the dynamics of the internal system is approximated to a unitary time evolution.
  We second derive the quantum Jarzynski equality based on the formulation of Hayashi and Tajima as a relation for the work measured by an external macroscopic apparatus.
  The right-hand side of the equality reduces to unity for ``natural'' cyclic processes, but fluctuates wildly for non-cyclic ones, exceeding unity often.
  This fluctuation should be detectable in experiments and provide evidence for the present formulation.
\end{abstract}


\maketitle

\section{Introduction}
  Thermodynamics was first introduced as a practical study to clarify the optimal performance of heat engines~\cite{Carnot1824} and has become one of the most important fields in physics~\cite{Fermi1956}.
  Recently, however, the development of experimental technology are realizing heat engines which are out of the scope of the standard thermodynamics, i.e., small-size heat engines.
  Quantum-scale and mesoscopic thermomotors, which used to be imaginary devices, are being realized in laboratory~\cite{Rousselet1994,Faucheux1995,Toyabe2010,An2014,Batalhao2014}.
  The functions of bio-molecules, which are micro-machines in nature, are being clarified too~\cite{Ishii2000}.

  We cannot apply the standard thermodynamics to these nanometer-size heat engines as it is, because it is a phenomenology for macroscopic systems.
  Statistical mechanics, another fundamental field of physics, has been applied to the small-size heat engines whose small-size working body is connected to the infinitely large heat baths, and is achieving a splendid success~\cite{Gallavotti1995,Jarzynski1997,Jarzynski1997a,Kurchan1998,Crooks1999,Seifert2005,Seifert2012,Ito2013,Kurchan2000,Tasaki2000,Bender2002,Talkner2007,Sagawa2008,Sagawa2009,Jacobs2009,Campisi2009,Campisi2011,Esposito2011,DeLiberato2011,Sagawa2012c,Funo2013,Tajima2013a,Tajima2013b,Tajima2014a,Funo2015}.

  Many studies of such engines~\cite{Kurchan2000,Tasaki2000,Bender2002,Talkner2007,Sagawa2008,Sagawa2009,Jacobs2009,Campisi2009,Campisi2011,Esposito2011,DeLiberato2011,Sagawa2012c,Funo2013,Tajima2013a,Tajima2013b,Tajima2014a,Funo2015} adopt a model of a microscopic internal quantum system connected to a macroscopic external agent.
  They use the following setup and assumption:
  \begin{enumerate}
    \item Thermodynamic operation of the internal quantum system by an external system is represented by a unitary operator of the time-dependent Hamiltonian of the internal quantum system which is controlled by external parameters. \label{enu:unitary_dyn}
    \item The work performed to the external system is equal to the energy loss of the internal quantum system.
  \end{enumerate}
  This approach has a practical advantage that we can formulate thermodynamic relations by analyzing only the internal quantum system. 

  However, there are always two concerns about the validity of this approach.
  One is about the assumption of the unitary dynamics and the other is about the definition of the work.
  In actual situations of heat engines, the internal system, whether quantum or not, is always attached to an external macroscopic system~\cite{Yamamoto2015}, and hence the dynamics of the internal quantum system cannot be exactly unitary. 
  It is not guaranteed either that all energy loss of the internal system becomes the work done to the external system.
  Is it accurate enough to approximate the true dynamics with a unitary dynamics?
  Is it legitimate to regard the energy loss as the extracted work?
  In spite of these concerns, the approach has been accepted by many researchers, because there have been results out of this approach which seem to be consistent with thermodynamics. 
  For example, we can derive some thermodynamic relations~\cite{Bender2002,Esposito2011,DeLiberato2011,Sagawa2012c} and a kind of quantum extension of the Jarzynski equality~\cite{Kurchan2000,Tasaki2000,Campisi2009,Campisi2011}.

  In the present paper, we face these two concerns.
  We claim here that the work out of a heat engine should be measured by means of a macroscopic system, for example, as movement of a macroscopic piston or wheel.
  From this point of view, the work extraction from the internal quantum system by the macroscopic system can be described as a standard quantum measurement process, in which the measurement result is evaluated on the side of the macroscopic system.
  In other words, the time evolution of the internal system is described as a quantum measurement process and the extracted work is regarded as a measurement outcome of this process. 
  This formulation, recently introduced by Hayashi and Tajima~\cite{Hayashi2015,TajimaThesis,Tajima2014}, raises a serious problem of the conventional approach in the form of a general trade-off relation.
  The relation shows that when the time evolution of the internal system is approximated to a unitary, it becomes difficult to fix the amount of extracted work.
  
  In the present paper, we first demonstrate the inappropriateness of the unitary time evolution of the internal system (namely the assumption~\ref{enu:unitary_dyn} above).
   It was pointed out by Hayashi and Tajima that the amount of the extracted work is not able to be evaluated on the side of the external system when the time evolution of the internal system is approximated to a unitary one~\cite{Hayashi2015}.
  It was also illustrated with a plain example by Tasaki, who is one of the advocates of the conventional approach~\cite{Tasaki2016}.
  In order to demonstrate this fact, we show for a simple model that the conventional approach has a problem regarding the fluctuation of the work.
  More specifically, we show that the variance of the energy transferred to the external system diverges when the time evolution of the internal system is approximated to a unitary, and is completely different from that of the energy loss of the internal system.
  This result demands us to change the derivation of the quantum Jarzynski equality from the previous ones~\cite{Kurchan2000,Tasaki2000,Campisi2009,Campisi2011} based on the conventional approach, which employed the \textit{internal} unitary time evolution and regarded the measured value of the energy loss of the internal system as the ``work'',
  whereas the energy gain of the \textit{external} system is the actual work that we can use.
  The Jarzynski equality derived from the previous approach does not contain relevant information about the fluctuation of the actual work.

  In order to resolve this problem, we next derive the quantum Jarzynski equality based on the measurement-based work extraction of Hayashi and Tajima~\cite{Hayashi2015,TajimaThesis}.
  This quantum Jarzynski equality is the relation of the work measured by the external macroscopic apparatus, which is equal to the measured value of the energy gain of the external system.
  Therefore, our derivation correctly contains the information about the fluctuation of the actual work, namely the energy gain of the external system.

  There has been another approach to the two concerns, in which they~\cite{Horodecki2013,Skrzypczyk2014} used an internal quantum system connected to an external quantum system.
  This approach assumes that the time evolution of the total system is unitary and the work extracted from the internal system is defined as the energy gain of the external system.
  We claim that the setup in our approach is more realistic than in this approach~\cite{Horodecki2013,Skrzypczyk2014} in the sense that our external system is a macroscopic measurement apparatus.

\section{Inappropriateness of the unitary time evolution \label{sec:two-level}}
  In this section, using a toy model, we consider the problem as to whether we can represent the time evolution of an internal system in terms of a unitary one.
  More specifically, we will show that the variance of the energy transferred to an external system $\Ext$ becomes large when we approximate the time evolution of the two-level system $\Inter$ by means of a unitary one.

  Let $H_\Inter$ and $H_\Ext$ denote the Hamiltonians of the internal system $\Inter$ and the external system $\Ext$, respectively:
  \begin{align}
    H_\Inter &:= \half{\Delta} \left( \sProj[\Inter]{1} - \sProj[\Inter]{0} \right) , \label{eq:two_Ham} \\
    H_\Ext &:= \sum_{n = -\infty}^{\infty} n \Delta \sProj[\Ext]{n} ,
  \end{align}
  where $\Delta$ is the level spacing, while $\sKet[\Ext]{n}$ with any integer $n$ are the energy eigenstates of $\Ext$ with the eigenvalue $n\Delta$.
  The fact that $n$ runs from negative infinity to positive infinity represents that the external system is macroscopic.

  We consider the case in which the time evolution operator $U_\IE$ of the total system is unitary and conserves the energy as in $\Commu{H_\Inter + H_\Ext}{U_\IE} = 0$.
  We can therefore decompose $U_\IE$ into the form
  \begin{equation}
    U_\IE := \sum_{n, n^\prime} K_{n, n^\prime} \otimes \sKetbra[\Ext]{n}{n^\prime}
    \label{eq:U_SE}
  \end{equation}
  with
  \begin{multline}
     K_{n, n^\prime} 
       := \delta_{n,n^\prime} \left( a_n \sProj[\Inter]{0} + b_n \sProj[\Inter]{1} \right) \\
       + \delta_{n+1, n^\prime} c_n \sKetbra[\Inter]{1}{0} + \delta_{n-1, n^\prime} d_n \sKetbra[\Inter]{0}{1}
    \label{eq:Kraus_op_K}
  \end{multline}
  for all integers $n$ and $n^\prime$, where the coefficients $a_n$, $b_n$, $c_n$ and $d_n$ are complex numbers and satisfy
  \begin{align}
    \Abs{a_n}^2 + \Abs{c_{n - 1}}^2 &= 1, \label{eq:a^2} \\
    \Abs{b_n}^2 + \Abs{d_{n + 1}}^2 &= 1, \label{eq:d^2} \\
    a_n d_n^\ast + b_{n-1}^\ast c_{n-1} &= 0  \label{eq:ad}
  \end{align}
  for all integers $n$.
  The time evolution of the internal system $\Inter$ is hence given by
  \begin{align}
    \Kmap{\rho_\Inter}
    &:= \bTr[\Ext]{U_\IE \left( \rho_\Inter \otimes \rho_\Ext \right) U_\IE^\dagger} \\
    &= \sum_{n, n^\prime, n^{\prime\prime}} \sBraket[\Ext]{n^\prime}{\rho_\Ext}{n^{\prime\prime}} K_{n, n^\prime} \rho_\Inter K_{n, n^{\prime\prime}}^\dagger ,
    \label{eq:time_evo_S}
  \end{align}
  where $\rho_\Inter$ and $\rho_\Ext$ are the initial states of $\Inter$ and $\Ext$, respectively, and $\Tr_\Ext$ denotes the trace operation with respect to the external system $\Ext$.
  The state $\rho_\Inter$ of $\Inter$ must be written in the form
  \begin{equation}
    \rho_\Inter
      := p \sProj[\Inter]{0} + \left( 1 - p \right)\sProj[\Inter]{1}
      + q^\ast \sKetbra[\Inter]{1}{0} + q \sKetbra[\Inter]{0}{1}
    \label{eq:arbitrary_density}
  \end{equation}
  with $0 \leq p \leq 1$, $q \in \mathbb{C}$.
  The positivity of $\rho_\Inter$ dictates the coefficient $q$ to satisfy $\Abs{q}^2 \leq p \left( 1 - p \right) \leq 1/4$.

  The probability that the energy 
  \begin{equation}
    w_j := h_j - \bTr{\rho_\Ext H_\Ext} \label{eq:two_lev_wj} 
  \end{equation}
  is transferred to the external system $\Ext$ during the time evolution is given by
  \begin{align}
    p_j &:= \bTr{\sProj[\Ext]{j} U_\IE \left( \rho_\Inter \otimes \rho_\Ext \right) U_\IE^\dagger} \\
        &= \sum_{n^\prime, n^{\prime\prime}} \sBraket[\Ext]{n^\prime}{\rho_\Ext}{n^{\prime\prime}} \bTr{K_{j, n^\prime} \rho_\Inter K_{j, n^{\prime\prime}}^\dagger } , \label{eq:prob_j}
  \end{align}
  where $h_j := j \Delta$ is an eigenvalue of the Hamiltonian $H_\Ext$ of the external system $\Ext$. 
  The energy conservation $\Commu{H_\Inter + H_\Ext}{U_\IE} = 0$ leads to
  \begin{widetext}
    \begin{align}
      \Avr[\Ext]{w} 
      &:= \sum_{j = -\infty}^{\infty} w_j p_j \\
      &= \bTr{H_\Ext U_\IE \left( \rho_\Inter \otimes \rho_\Ext \right) U_\IE^\dagger} - \bTr{\rho_\Ext H_\Ext} \\
      &= \bTr{\left( H_\Inter + H_\Ext \right) \left( \rho_\Inter \otimes \rho_\Ext \right)} -\bTr{H_\Inter U_\IE \left( \rho_\Inter \otimes \rho_\Ext \right) U_\IE^\dagger} - \bTr{\rho_\Ext H_\Ext} \\
      &= \bTr{\rho_\Inter H_\Inter} - \bTr{\Kmap{\rho_\Inter} H_\Inter} .
      \label{eq:avr_w_S}
    \end{align}
  \end{widetext}
  The variance $V\left( w \right)$ of the energy transfer is given by
  \begin{equation}
    V_\Ext\left( w \right)
    := \Avr[\Ext]{w^2} - \Avr[\Ext]{w}^2 
    = \Avr[\Ext]{h^2} - \Avr[\Ext]{h}^2,
  \end{equation}
  where $\Avr[\Ext]{h}$ and $\Avr[\Ext]{h^2}$ are the average and the mean square of $h_j$, respectively, with respect to $p_j$.

  So far, everything has been exact.
  Now, we try to approximate the time evolution of $\Inter$ to a specific unitary matrix
  \begin{equation}
    U_\Inter := \sum_{k, l = 0, 1} u_{k, l} \sKetbra[\Inter]{k}{l} .
  \end{equation}
  Here, the approximation to the unitary matrix $U_\Inter$ means the following condition: for any $\epsilon > 0$, the inequality $\dist{\Kmap{\rho_\Inter}}{U_\Inter \rho_\Inter U_\Inter^\dagger} < \epsilon$ holds for all state $\rho_\Inter$, where $d$ is a distance function.

  Using the example given in Ref.~\cite{Aberg2014,TajimaThesis}, we now choose the initial state of the external system $\Ext$ and the coefficients of Eq.~\eqref{eq:Kraus_op_K} in the form
  \begin{equation}
    \rho_\Ext := \sProj[\Ext]{\psi}, \quad 
    \sKet[\Ext]{\psi} := \inv{\sqrt{M}} \sum_{m = 1}^{M} \sKet[\Ext]{m} ,
    \label{eq:init_psi} 
  \end{equation}
  \begin{multline}
    K_{n, n^\prime}
      = \delta_{n,n^\prime} \left( u_{0,0} \sProj[\Inter]{0} + u_{1,1} \sProj[\Inter]{1} \right) \\
      + \delta_{n+1, n^\prime} u_{1,0} \sKetbra[\Inter]{1}{0} + \delta_{n-1, n^\prime} u_{0,1} \sKetbra[\Inter]{0}{1}
      \label{eq:approx_K}
  \end{multline}
  with $M := 8\left\lceil \epsilon^{-2} \right\rceil$ and all integrals $n$ and $n^\prime$. 
  Then, we obtain~\cite{TajimaThesis}
  \begin{equation}
    b\left( \Kmap{\rho_\Inter}, U_\Inter \rho_\Inter U_\Inter^\dagger \right) < \epsilon,
  \end{equation}
  where $b$ is the Bures distance.
  Because the initial state is far from an energy eigenstate, we can indeed approximate the time evolution of $\Inter$ to $U_\Inter$.

  A problem arises as follows, however.
  Combining Eqs.~\eqref{eq:U_SE} and \eqref{eq:init_psi}, we obtain Eq.~\eqref{eq:prob_j} in the form
  \begin{equation}
    p_j = \inv{M} \bTr{\bar{K}_j^\dagger \bar{K}_j \rho_\Inter} 
  \end{equation}
  with 
  \begin{equation}
    \bar{K}_j := \sum_{m = 1}^M K_{j, m}.
    \label{eq:K_bar}
  \end{equation}
  The expectation and the mean square of $h_j$ are respectively given by
  \begin{align}
    \Avr[\Ext]{h} &= \sum_j h_j p_j 
    = \frac{\Delta}{M} \sum_j j \bTr{\bar{K}_j^\dagger \bar{K}_j \rho_\Inter} , \label{eq:expect_h} \\
    \Avr[\Ext]{h^2} &= \sum_j \left( h_j \right)^2 p_j 
    = \frac{\Delta^2}{M} \sum_j  j^2 \bTr{\bar{K}_j^\dagger \bar{K}_j \rho_\Inter}  . \label{eq:expect_h2}
  \end{align}
  After the algebra in App.~\ref{app:app_B}, we obtain 
  \begin{equation}
    \Avr[\Ext]{h} = \half{\Delta} \left( M + 1 \right) + \Avr[\Ext]{w} , \label{eq:avr_h} 
  \end{equation}
  \begin{multline}
    \Avr[\Ext]{h^2} 
      =  \frac{\Delta^2}{6} \left( M + 1 \right)\left( 2M + 1 \right) + \Delta \left( M + 1 \right) \Avr[\Ext]{w} \\
      + \Delta^2 \left( \Abs{u_{0,1}}^2 \left( 1 - p \right) + \Abs{u_{1,0}}^2 p \right) ,
      \label{eq:avr_h_2}  
  \end{multline}
  where $\Avr[\Ext]{w}$ is the expectation of $w_j$ in the form
  \begin{multline}
    \Avr[\Ext]{w} = \Delta \Bigl[ \Abs{u_{0,1}}^2 \left( 1 - p \right) - \Abs{u_{1,0}}^2 p  \\
      + 2\left( 1 - M^{-1} \right) \text{Re}\left( u_{0,0} u_{0,1}^\ast q \right) \Bigr] .
  \end{multline}
  Therefore, we obtain
  \begin{multline}
    V_\Ext\left( w \right)
      = \frac{\Delta^2}{12} \left( M^2 - 1 \right) - \Avr[\Ext]{w}^2 \\
      + \Delta^2 \left( \Abs{u_{0,1}}^2 \left( 1 - p \right) + \Abs{u_{1,0}}^2 p \right).
    \label{eq:variance_E}
  \end{multline}
  Because $\Avr[\Ext]{w} = \Order{1}$, the variance of the energy transfer diverges as $M = 8\left\lceil \epsilon^{-2} \right\rceil \to \infty$ in the limit $\epsilon \to 0$.
  In other words, we cannot fix the amount of the energy transfer when we approximate the dynamics of $\Inter$ by a unitary one.
  This demonstrates that it is not appropriate to use the unitary dynamics for the internal system $\Inter$ and define the work as the energy loss of $\Inter$.

  Let us compare this with the variance of the energy loss of the internal system $\Inter$.
  We measure the energy of the internal system $\Inter$ before and after the time evolution~\eqref{eq:time_evo_S} and regard the difference of the two measurement outcomes as the energy loss of the internal system, which the previous derivation of the quantum Jarzynski equalities~\cite{Kurchan2000,Tasaki2000,Campisi2009,Campisi2011} defined as the work.
  Because the time evolution of the internal system $\Inter$ is given by Eq.~\eqref{eq:time_evo_S}, the probability of the energy loss of the internal system
  \begin{equation}
    \upsilon_{\alpha, \beta} := e_\alpha - e_\beta 
  \end{equation}
  is given by
  \begin{equation}
    q_{\alpha, \beta} := \sBraket[\Inter]{\alpha}{\rho_\Inter}{\alpha} \bTr{\sProj[\Inter]{\beta} \Kmap{\sProj[\Inter]{\alpha}} } ,
    \label{eq:inter_prob}
  \end{equation}
  where $e_0 := -\Delta / 2$ and $e_1 := \Delta / 2$ are the eigenvalues of the Hamiltonian $H_\Inter$.
  The variance $V_\Inter\left( \upsilon \right)$ of the energy loss of the internal system is hence given by
  \begin{equation}
    V_\Inter\left( \upsilon \right)
    := \Avr[\Inter]{\upsilon^2} - \Avr[\Inter]{\upsilon}^2 ,
  \end{equation}
  where $\Avr[\Inter]{\upsilon}$ and $\Avr[\Inter]{\upsilon^2}$ are the average and the mean square of $\upsilon_{\alpha, \beta}$, respectively, with respect to $q_{\alpha, \beta}$.
  Since $e_0 := -\Delta / 2$ and $e_1 := \Delta / 2$, the energy loss $\upsilon_{\alpha, \beta}$ is $-\Delta$, $0$ or $\Delta$.
  Therefore, we obtain
  \begin{equation}
    \Avr[\Inter]{\upsilon^2}
    := \sum_{\alpha, \beta = 0, 1} q_{\alpha, \beta} \upsilon_{\alpha, \beta}^2
      \leq \Delta^2 ,
  \end{equation}
  and hence
  \begin{equation}
    V_\Inter\left( \upsilon \right)
    := \Avr[\Inter]{\upsilon^2} - \Avr[\Inter]{\upsilon}^2 
    \leq \Delta^2 .
  \end{equation}
  We thus see that $V_\Inter(\upsilon)$ appears to be completely different from $V_\Ext(w)$ in Eq.~\eqref{eq:variance_E} when we approximate the time evolution of the internal system to a unitary one.

  The above demonstration raises a problem of the quantum Jarzynski equalities derived in the previous approach~\cite{Kurchan2000,Tasaki2000,Campisi2009,Campisi2011}, which employed a unitary for the time evolution of the internal system and regarded the energy loss of the internal system as the work.
  Because the energy gain of the external system is the actual work that we can use, the work defined in the previous approach as well as the Jarzynski equalities derived thereby do not contain relevant information about the fluctuation of the actual work.
  In order to resolve this problem, we derive in Sec.~\ref{sec:JE} the quantum Jarzynski equality using the measurement-based work extraction of Hayashi and Tajima~\cite{Hayashi2015,TajimaThesis}.

  Incidentally, in the case of $M = 1$, we cannot approximate the time evolution to a unitary one.
  To show it, we consider the quantity
  \begin{equation}
    \min_{U_\Inter : \text{unitary}} \dist[\Tr]{\Kmap{\rho_\Inter} }{U_\Inter \rho_\Inter U_\Inter^\dagger} ,
    \label{eq:min_distance}
  \end{equation}
  where $d_{\Tr}$ is the trace distance.
  For $M = 1$, the initial state~\eqref{eq:init_psi} of $\Ext$ is a pure energy eigenstate with a fixed energy level $n_0$, namely, $\rho_\Ext = \sProj[\Ext]{n_0}$.
  Then, Eq.~\eqref{eq:time_evo_S} reduces to
  \begin{widetext}
    \begin{multline}
      \Kmap{\rho_\Inter}
        = \sum_n K_{n, n_0} \rho_\Inter K_{n, n_0}^\dagger \\
        = \left[ \Abs{a_{n_0}}^2 p + \Abs{d_{n_0 + 1}}^2 \left( 1 - p \right) \right] \sProj[\Inter]{0} 
          + \left[ \Abs{b_{n_0}}^2 \left( 1 - p \right) + \Abs{c_{n_0 - 1}}^2 p \right] \sProj[\Inter]{1}
          + a_{n_0}^\ast b_{n_0} q^\ast \sKetbra[\Inter]{1}{0}
          + a_{n_0} b_{n_0}^\ast q \sKetbra[\Inter]{0}{1} .
      \label{eq:pure_time_evo_S}
    \end{multline}
  \end{widetext}
  From Eq.~\eqref{eq:avr_w_S}, we have the average of the transferred energy in the form
  \begin{equation}
    \Avr[\Ext]{w} = \Delta \left[ \Abs{d_{n_0 + 1}}^2 \left( 1 - p \right) - \Abs{c_{n_0 - 1}}^2 p\right].
    \label{eq:pure_avr_w}
  \end{equation}

  Since the states $\Kmap{\rho_\Inter}$ and $\rho_\Inter$ are Hermitian operators, we can diagonalize them using unitary operators.
  In other words, there exist unitary operators $V$ and $V^\prime$ such that
  \begin{align}
    \rho_\Inter &= V \Lambda V^\dagger , \label{eq:diag_rhoS} \\
    \Kmap{\rho_\Inter} &= V^\prime \Lambda^\prime {V^\prime}^\dagger , \label{eq:diag_K_rhoS}
  \end{align}
  where $\Lambda$ and $\Lambda^\prime$ are diagonal matrices.
  Because of the unitary invariance of the trace distance, Eq.~\eqref{eq:min_distance} becomes
  \begin{equation}
    \min_{U_\Inter : \text{unitary}} \dist[\Tr]{\Kmap{\rho_\Inter}}{U_\Inter \rho_\Inter U_\Inter^\dagger}
    = \min_{\tilde{U}_\Inter : \text{unitary}} \dist[\Tr]{\tilde{U}_\Inter \Lambda^\prime \tilde{U}_\Inter^\dagger}{\Lambda} .
  \end{equation}
  The calculation in App.~\ref{app:app_A} then gives
  \begin{align}
    \min_{\tilde{U}_\Inter : \text{unitary}} \dist[\Tr]{\tilde{U}_\Inter \Lambda^\prime \tilde{U}_\Inter^\dagger}{\Lambda} 
    &= \half{1} \min_{\tilde{U}_\Inter : \text{unitary}} \Tr \Abs{ \tilde{U}_\Inter \Lambda^\prime \tilde{U}_\Inter^\dagger - \Lambda} \\
    &= \Abs{R^\prime - R} 
    \label{eq:d_R-R}
  \end{align}
  with
  \begin{gather}
    R := \sqrt{\left( p - \half{1} \right)^2 + \Abs{q}^2} , \label{eq:R} \\
    R^\prime := \sqrt{\left( \left( 1 - x - y \right)p + x - \half{1} \right)^2 + \left( 1 - x \right) \left( 1 - y \right) \Abs{q}^2} , \label{eq:R^prime} \\
    x := \Abs{d_{n_0 + 1}}^2, \quad y := \Abs{c_{n_0 - 1}}^2 .
  \end{gather}

  \begin{figure*}[tbp]
    \begin{tabular}{cc}
      \includegraphics[width=0.47\hsize]{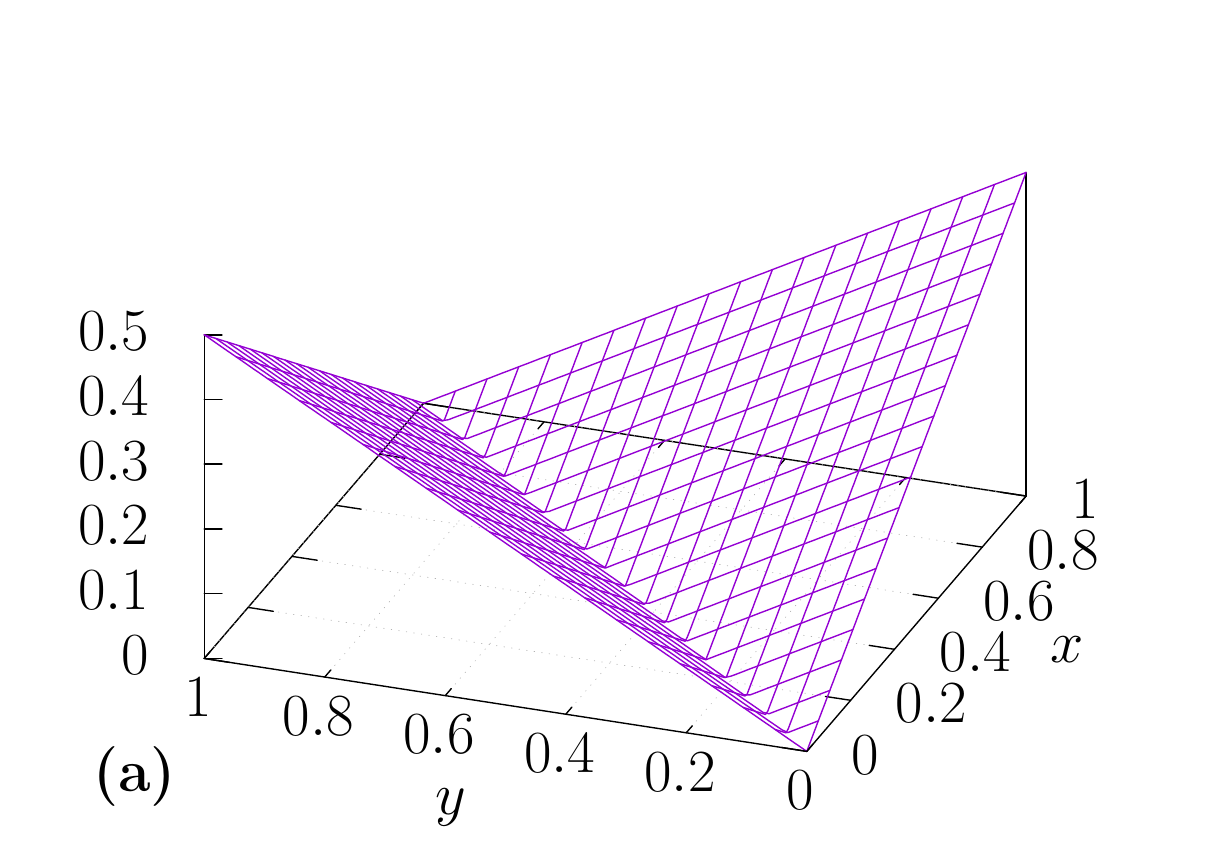} &
      \includegraphics[width=0.47\hsize]{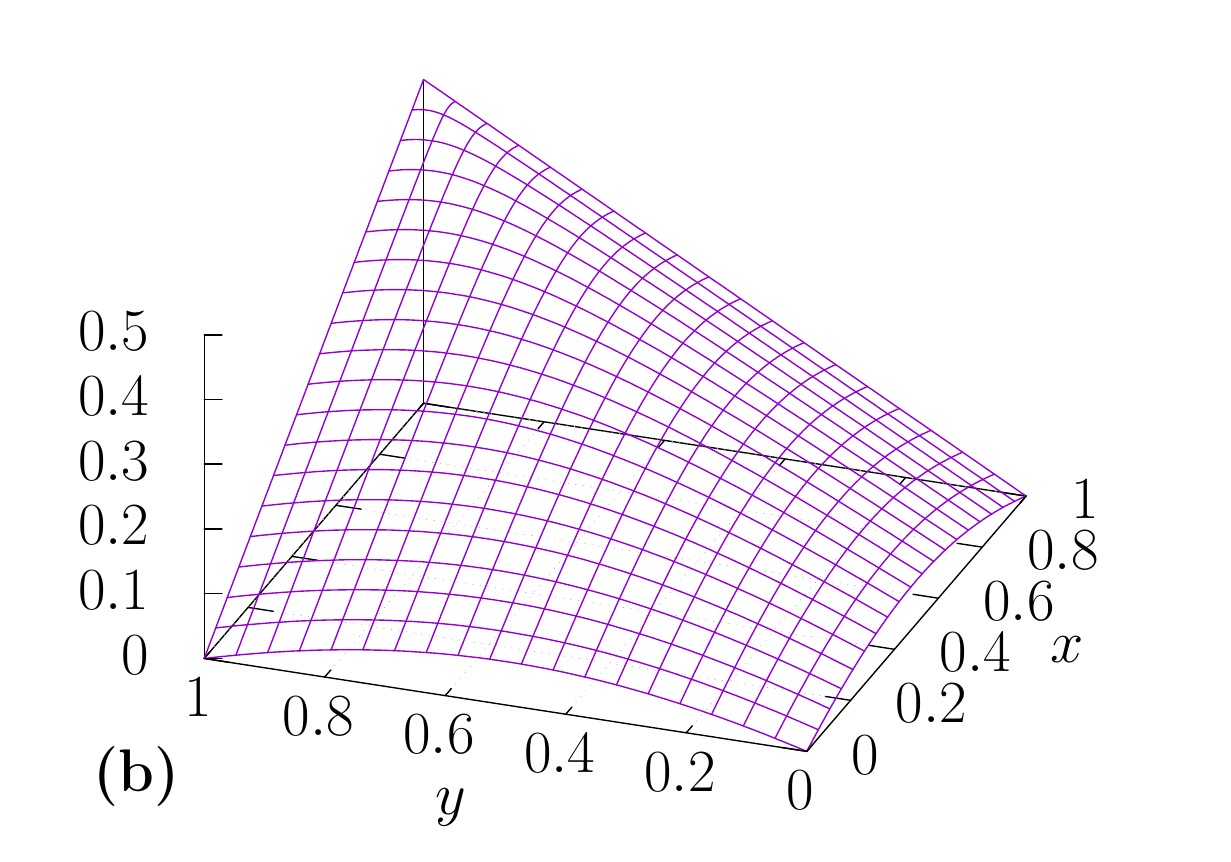} \\
    \end{tabular}
    \caption{(Color online) Plots of Eq.~\eqref{eq:d_R-R}. We take the parameters (a)~$\left( p, \Abs{q} \right) = \left( 1/2, 0 \right)$ and (b)~$\left( 1/2, 1/2 \right)$. \label{fig:R'-R}}
  \end{figure*}
  In the case of $x = y = 0$, the distance of Eq.~\eqref{eq:d_R-R} is equal to zero.  However, the energy transfer~\eqref{eq:pure_avr_w} is trivially equal to zero in this case.
  For that reason, we consider cases other than $x = y = 0$.
  When we choose the initial state of $\Inter$ as $p = 1 / 2$ and $\Abs{q} = 0$, the distance of Eq.~\eqref{eq:d_R-R} is not equal to zero (Fig.~\ref{fig:R'-R}(a)).
  When, we choose the initial state of $\Inter$ as $p = 1 / 2$ and $\Abs{q} = 1/2$, the distance of Eq.~\eqref{eq:d_R-R} is also not equal to zero for $x = y \neq 0$ (Fig.~\ref{fig:R'-R}(b)).
  Consequently, we cannot approximate the time evolution to a unitary when $M = 1$.
  
\section{Jarzynski equality \label{sec:JE}}
  As the main result of the present paper, we here derive the Jarzynski equality for the measurement-based work extraction.
  Previous quantum versions of the Jarzynski equalities~\cite{Kurchan2000,Tasaki2000,Campisi2009,Campisi2011} assumed that the time evolution of the internal system was a unitary one and the work performed to the external system was equal to the energy loss of the internal system.
  However, as shown in Sec.~\ref{sec:two-level}, this is not appropriate as the definition of the work.
  For this reason, we introduce a new derivation of the quantum Jarzynski equality based on the formulation of Hayashi and Tajima~\cite{Hayashi2015,TajimaThesis}.

\subsection{Cyclic process \label{sec:closed}}
  We first consider a cyclic process, in which the final Hamiltonian is equal to the initial one.
  The external system $\Ext$ receives energy from the internal system $\Inter$ between time $t = 0$ to $T$.
  The time evolution of the total system between time $t = 0$ to $T$ is unitary.

  Let $H_\Inter$ and $H_\Ext$ denote the time-independent Hamiltonians of $\Inter$ and $\Ext$, respectively, and $H_\inter(t)$ denote the time-dependent Hamiltonian interacting between $\Inter$ and $\Ext$; the total system is given by
  \begin{equation}
    H_\tot(t) := H_\Inter + H_\Ext + H_\inter(t).
  \end{equation}
  The eigenvalue decompositions of $H_\Inter$ and $H_\Ext$ are denoted by
  \begin{align}
    H_\Inter &:= \sum_x h_x \sProj[\Inter]{h_x} , \label{eq:cl_S_Ham} \\
    H_\Ext &:= \sum_i e_i \sProj[\Ext]{e_i} , \label{eq:E_Ham}
  \end{align}
  where $h_x$ and $e_i$ are the eigenvalues of $H_\Inter$ and $H_\Ext$, respectively.
  We assume that the interaction Hamiltonian satisfies the condition
  \begin{equation}
    \Commu{H_\Inter + H_\Ext}{U_\IE} = 0 ,
    \label{eq:conserv_energy}
  \end{equation}
  where $U_\IE$ is the time evolution of total system given by~\cite{Hayashi2015}
  \begin{equation}
    U_\IE := \mathcal{T} \exp\left( -\frac{i}{\hbar} \int_0^T H_\tot(\tau) d\tau \right) ,
  \end{equation}
  where $\mathcal{T}$ is time-ordered product.
  The condition of Eq.~\eqref{eq:conserv_energy} means that the total energy from $H_\Inter + H_\Ext$ is the same before and after $U_\IE$, and hence the net energy from the interaction Hamiltonian is equal to zero.
  We further assume that the initial states of $\Inter$ and $\Ext$ are the canonical distribution at an inverse temperature $\beta$ and a pure eigenstate of energy $e_0$, respectively:
  \begin{gather}
    \rho_{\Inter, \can} := \sum_x \frac{e^{-\beta h_x}}{Z_\Inter} \sProj[\Inter]{h_x} , \label{eq:in_init_S} \\
    \rho_\Ext := \sProj[\Ext]{e_0} \label{eq:in_init_E}
  \end{gather}
  with $Z_\Inter := \bTr{e^{-\beta H_\Inter}}$.
  
  We then consider the following process:
  \begin{enumerate}
    \item We set the initial states given by Eqs.~\eqref{eq:in_init_S} and \eqref{eq:in_init_E}.
    \item We then let the total system evolve under the unitary operator $U_\IE$.
          The key here is to consider the unitary time evolution of the total system, not of the internal system $\Inter$.
    \item We finally measure the energy of the external system $\Ext$ using the projection operator $\sProj[\Ext]{e_j}$ and define $\Delta e_j := e_j - e_0$ as the energy gain.
          It is essential that at this point the ``work'' $\Delta e_j$ is not a fixed value but given probabilistically.
          \label{enu:cyc_process_3}
  \end{enumerate}
  In the above process, the time evolution of $\Inter$ and the measurement process of the specific energy gain $\Delta e_j$ are defined as
  \begin{equation}
    \Emap{\rho_\Inter} := \bTr[\Ext]{U_\IE \left( \rho_\Inter \otimes \sProj[\Ext]{e_0} \right) U_\IE^\dagger}
    \label{eq:evo_I} 
  \end{equation}
  with $\calE = \sum_j \calE_j$ for all density operators $\rho_\Inter$ of $\Inter$, where
  \begin{equation}
    \Emap[_j]{\rho_\Inter} := \bTr[\Ext]{ \sProj[\Ext]{e_j} U_\IE \left( \rho_\Inter \otimes \sProj[\Ext]{e_0} \right) U_\IE^\dagger} .
    \label{eq:meas_proc}
  \end{equation}
 
  We now introduce the probability distribution of the extracted work $W$ as
  \begin{equation}
    P\left( W \right) := \sum_j \delta\left( W - \Delta e_j \right) \bTr{\Emap[_j]{\rho_{\Inter, \can}}} ,
  \end{equation}
  where $\delta\left( x \right)$ is the delta function.
  Because $\calE_j$ is a linear operator, we have
  \begin{multline}
    P\left( W \right)
    = \sum_{j, x, y} \delta\left( W - \Delta e_j \right) \\
    \times \frac{e^{-\beta h_x}}{Z_\Inter} \bTr{\sProj[\Inter]{h_y} \Emap[_j]{\sProj[\Inter]{h_x}} },  
  \end{multline}
  where we inserted a resolution of unity $\sum_y \sProj[\Inter]{h_y}$.
  Since the total energy $H_\Inter + H_\Ext$ does not change after $U_\IE$ and the external system gains the energy $\Delta e_j$ after the process $\calE_j$, the energy of the internal system must be $h_y = h_x - \Delta e_j$ outside the operator $\calE_j$.
  We therefore find
  \begin{widetext}
    \begin{align}
      P\left( W \right)
      &= e^{-\beta W} \sum_{j, x, y} \delta\left( W - \Delta e_j \right) \frac{e^{-\beta h_y}}{Z_\Inter} \bTr{\sProj[\Inter]{h_y} \Emap[_j]{\sProj[\Inter]{h_x}} } \\
      &= e^{-\beta W} \sum_j \delta\left( W - \Delta e_j \right) \bTr{ \rho_{\Inter, \can} \Emap[_j]{\id_\Inter} } . \label{eq:cal_P(W)}
    \end{align}
  \end{widetext}
  Let us denote the average with respect to $P\left( W \right)$ by
  \begin{equation}
    \Avr{f(W)} := \int dW\, f(W) P(W) , \label{eq:def_avr}
  \end{equation}
  where $f(W)$ is an arbitrary function of the extracted work $W$.
  From Eq.~\eqref{eq:cal_P(W)}, we can therefore obtain the Jarzynski equality under the cyclic process in the form
  \begin{equation}
    \Avr{e^{\beta W}} = \gamma
    \label{eq:Jarzyn_eq}
  \end{equation}
  with 
  \begin{equation}
    \gamma := \bTr{\Emap[^\dagger]{\rho_{\Inter, \can}} },
  \end{equation}
  where $\calE^\dagger$ is the adjoint map of $\calE$, given by $\bTr{\Emap[^\dagger]{A^\dagger} B} = \bTr{A^\dagger \Emap{B}}$.
  Note that we did not use the details of the external system $\Ext$ but the energy conservation of the time evolution, Eq.~\eqref{eq:conserv_energy}.

  Applying Jensen's inequality $\Avr{e^f} \geq e^{\Avr{f}}$ to Eq.~\eqref{eq:Jarzyn_eq}, we obtain
  \begin{equation}
    \Avr{W} \leq \beta^{-1} \log \gamma .
    \label{eq:ex_Kelvin_prin}
  \end{equation}
  This inequality is the second law of thermodynamics under the measurement-based work extraction.

  In the high-temperature limit $\beta = 0$, the initial state~\eqref{eq:in_init_S} reduces to $\rho_{\Inter, \can} = \id_\Inter / \mathcal{N}$, where $\mathcal{N}$ is the dimensionality of the Hilbert space of the internal system $\Inter$.
  Thus, owing to the linearity and the trace preserving of $\calE$, the quantity $\gamma$ reduces to unity.

  For general values of $\beta$, let us assume that the time evolution of $\Inter$ is a ``natural'' thermodynamic process; that is, the measurement process is not a feedback process and satisfies the first and second laws of thermodynamics for an arbitrary initial state.
  Hayashi and Tajima introduced and called it the standard CP-work extraction~\cite{Hayashi2015,TajimaThesis}.
  The first law is satisfied by the measurement process $\calE_j$ which changes the energy eigenstate $\sKet[\Inter]{h_x}$ of $\Inter$ to the state of the energy $h_x - \Delta e_j$ when the external system gains the energy $\Delta e_j$.
  The second law corresponds to the time evolution $\calE = \sum_j \calE_j$ which satisfies 
  \begin{equation}
    S\left( \rho_\Inter \right) \leq S\left( \Emap{\rho_\Inter} \right) 
    \label{eq:unital_cond}
  \end{equation}
  for all initial states $\rho_\Inter$ of $\Inter$, where $S\left( \rho_\Inter \right) := -\bTr{\rho_\Inter \log \rho_\Inter}$ is the von Neumann entropy.
  As a necessary and sufficient condition of \eqref{eq:unital_cond} for all initial states, the time evolution $\calE$ must be a unital map~\cite{Kimura2009}; 
  \begin{equation}
    \Emap{\id_\Inter} = \id_\Inter .
    \label{eq:unital}
  \end{equation}
  Then, the quantity $\gamma$ is unity, and Eqs.~\eqref{eq:Jarzyn_eq} and \eqref{eq:ex_Kelvin_prin} reduce to
  \begin{gather}
    \Avr{e^{\beta W}} = 1 , \\
    \Avr{W} \leq 0 , \label{eq:Kelvin_prin}
  \end{gather}
  respectively.
  Hence, we obtain the same form as the previous derivations~\cite{Kurchan2000,Tasaki2000,Campisi2009,Campisi2011} of the Jarzynski equality under the cyclic process.

  The difference of the Jarzynski equality from unity is known for feedback processes~\cite{Sagawa2010,Morikuni2011,Funo2013} and/or absolutely irreversible processes~\cite{Murashita2014,Funo2015}.
  Because the time evolution $\calE$ includes the feedback process, Eq.~\eqref{eq:Jarzyn_eq} also applies to the feedback process, for which the quantity $\gamma$ denotes the efficiency~\cite{Sagawa2010}.
  On the other hand, because we assume that the initial state~\eqref{eq:in_init_S} of the internal system is the canonical distribution, Eq.~\eqref{eq:Jarzyn_eq} does not apply to the absolutely irreversible process.

  We stress, however, that the present result is essentially different from the previous ones~\cite{Kurchan2000,Tasaki2000,Campisi2009,Campisi2011}.
  The previous derivation of the Jarzynski equality did not contain information about the fluctuation of the energy gain of the \textit{external} system appropriately, defining the measured value of the energy loss of the \textit{unitarily evolving internal} system as the random variable $W$.
  As we have shown in Sec.~\ref{sec:two-level}, however, under the approximation of the unitary dynamics of the internal system, the variance of $W$ in the previous derivation is completely different from that of the energy gain of the external system, which is the actual work that we can use.
  The Jarzynski equality derived from the previous formulation therefore does not give relevant information about the fluctuation of the actual work.

  Our derivation of the Jarzynski equality is different in this point.
  We also define the measured value of the energy loss of the internal system as a random variable $W$, but it is equal to the measured value of the energy gain of the external system, because now we employ the unitary time evolution of the total system satisfying Eq.~\eqref{eq:conserv_energy}.
  Therefore, our derivation correctly contains the information about the fluctuation of the actual work, namely the energy gain of the external system.
  For a possible extension of the present formulation to the measurement process with error, see Sec.~\ref{sec:Conclustion}.

\subsection{non-cyclic process \label{sec:opened}}
  Next, we consider the Jarzynski equality under a non-cyclic process, extending the case of the cyclic process in Sec.~\ref{sec:closed}.
  For a non-cyclic process, the energy spectrum of the internal system is different between the initial and final Hamiltonians.
  To apply the formalism for the cyclic process to the non-cyclic one, we divide the internal system $\Inter$ into two subsystems, namely a (further) internal system $\Sys$ and a control system $\Con$~\cite{Horodecki2013,TajimaThesis}.
  The internal system $\Sys$ is a working substance, such as a gas, while the control system $\Con$ controls the Hamiltonian of the internal system $\Sys$ as a piston.
  We consider the work extracted from the internal system $\Sys$.

  We assume the initial Hamiltonian~\eqref{eq:cl_S_Ham} of the internal system $\Inter$ in the form
  \begin{align}
    H_\Inter 
    &:= \sum_\lambda H_\Sys\left( \lambda \right) \otimes \sProj[\Con]{\lambda} \notag \\
    &= \sum_\lambda \sum_{\xbk{\lambda}} h_{\xbk{\lambda}} \sProj[\SC]{h_{\xbk{\lambda}}, \lambda},
    \label{eq:SC_Ham}
  \end{align}
  where $H_\Sys\left( \lambda \right)$ is the Hamiltonian of $\Sys$, whose eigenstate and the corresponding eigenvalue are denoted as $\sKet[\Sys]{h_{\xbk{\lambda}}}$ and $h_{\xbk{\lambda}}$, respectively, $\set{\sKet[\Con]{\lambda}}$ is an orthonormal basis of the control system $\Con$, and $\sKet[\SC]{h_{\xbk{\lambda}}, \lambda}$ denotes $\sKet[\Sys]{h_{\xbk{\lambda}}} \otimes \sKet[\Con]{\lambda}$.
  We vary the control parameter $\lambda$, making the process non-cyclic.
  Note that we made the $\lambda$ dependence of the index $\xbk{ \lambda }$ explicit, because the set of the eigenvalues of $H_\Sys\left( \lambda \right)$ depends on $\lambda$.
  The energy of $\Sys$ changes from $h_{\xbk{\lambda}}$ to $h_{\xbk{\lambda}} - \Delta e_j$ after the measurement process $\calE_j$ of the specific energy gain $\Delta e_j$.

  We set the initial state of $\Inter$ to be the canonical distribution of $\Sys$ with a pure state $\sKet[\Con]{\lambda_\init}$ of $\Con$:
  \begin{align}
    \rho_{\Inter}\left( \lambda_\init \right)
    &:= \sum_{\xbk{\lambda_\init}} \frac{e^{-\beta h_{\xbk{\lambda_\init}}}}{Z_\Sys\left(\lambda_\init\right)} \sProj[\SC]{h_{\xbk{\lambda_\init}}, \lambda_\init} \notag \\
    &= \rho_{\Sys, \can}\left( \lambda_\init \right) \otimes \sProj[\Con]{\lambda_\init} 
    \label{eq:can_IC}
  \end{align}
  with $Z_\Sys\left(\lambda\right) := \bTr{e^{-\beta H_\Sys\left( \lambda \right)}}$ and $\rho_{\Sys, \can}\left( \lambda \right) := e^{-\beta H_\Sys\left( \lambda \right)} / Z_\Sys\left(\lambda\right)$.
  This means that the internal system $\Sys$ starts from the equilibrium with the fixed parameter $\lambda_\init$.
  The free energy of $\Sys$ for a specific value of $\lambda$ is given by
  \begin{equation}
    F_\Sys\left( \lambda \right) := -\beta^{-1} \log Z_\Sys\left( \lambda \right) .
  \end{equation}

  We define the probability distribution of the extracted work $W$ during the process in which the state of $\Con$ changes from $\lambda_\init$ to $\lambda_\fin$ as
  \begin{multline}
    P_{\lambda_\init \to \lambda_\fin}\left( W \right)
    := \sum_j \frac{\delta\left( W - \Delta e_j \right)}{p_{\lambda_\init \to \lambda_\fin}} \\
    \times \bTr{\sProj[\Con]{\lambda_\fin} \Emap[_j]{\rho_{\Sys, \can}\left( \lambda_\init \right) \otimes \sProj[\Con]{\lambda_\init}}} ,
    \label{eq:open_P}
  \end{multline}
  where
  \begin{equation}
    p_{\lambda_\init \to \lambda_\fin}
    := \bTr{\sProj[\Con]{\lambda_\fin} \Emap{\rho_{\Sys, \can}\left( \lambda_\init \right) \otimes \sProj[\Con]{\lambda_\init}}}
    \label{eq:trans_prob}
  \end{equation}
  is the transition probability that the state of $\Con$ changes from $\lambda_\init$ to $\lambda_\fin$.
  In the same way as in Eq.~\eqref{eq:cal_P(W)} of Sec.~\ref{sec:closed}, we obtain
  \begin{widetext}
    \begin{equation}
      P_{\lambda_\init \to \lambda_\fin}\left( W \right)
        = e^{-\beta W} e^{-\beta \Delta F_\Sys\left( \lambda_\init, \lambda_\fin \right)} 
        \sum_j \frac{\delta\left( W - \Delta e_j \right)}{p_{\lambda_\init \to \lambda_\fin}}
        \bTr{\rho_{\Sys, \can}\left( \lambda_\fin \right) \otimes \sProj[\Con]{\lambda_\fin} \Emap[_j]{\id_\Sys\left( \lambda_\init \right) \otimes \sProj[\Con]{\lambda_\init}}} 
      \label{eq:open_P_after}
    \end{equation}
  \end{widetext}
  with $\Delta F_\Sys\left( \lambda_\init, \lambda_\fin \right) := F_\Sys\left( \lambda_\fin \right) - F_\Sys\left( \lambda_\init \right)$, while $\id_\Sys\left( \lambda_\init \right) := \sum_{\xbk{\lambda_\init}} \sProj[\Sys]{h_{\xbk{\lambda_\init}}}$ is the identity operator of $\Sys$ with fixed parameter $\lambda_\init$.

  We modify the average~\eqref{eq:def_avr} to
  \begin{equation}
    \Avr[\lambda_\init \to \lambda_\fin]{f(W)}
    := \int dW\, f(W) P_{\lambda_\init \to \lambda_\fin}(W) .
  \end{equation}
  We therefore arrive at the Jarzynski equality under a non-cyclic process in the form
  \begin{equation}
    \Avr[\lambda_\init \to \lambda_\fin]{e^{\beta W}}
    = \gamma_{\lambda_\init \to \lambda_\fin} e^{-\beta \Delta F_\Sys\left(\lambda_\init, \lambda_\fin\right)}  
    \label{eq:opened_Jar}
  \end{equation}
  with 
  \begin{gather}
    \gamma_{\lambda_\init \to \lambda_\fin}
      := \frac{q_{\lambda_\fin \to \lambda_\init}}{p_{\lambda_\init \to \lambda_\fin}} , \label{eq:gamma_2} \\
    q_{\lambda_\fin \to \lambda_\init}
      \!:= \bTr{\rho_{\Sys, \can}\!\left( \lambda_\fin \right) \!\otimes\! \sProj[\Con]{\lambda_\fin} \Emap{\id_\Sys\left( \lambda_\init \right) \otimes \sProj[\Con]{\lambda_\init}}} . \label{eq:qusi_trans_prob}
  \end{gather}
  We note that the state of $\Con$ is measured only in the initial and final states.
  During the dynamics between these states, we cannot tell the path of the change of physical quantities of $\Con$, such as the position of a piston, nor can we the motion of $\Sys$.
  It is in contrast with the fact that in the previous derivation of the Jarzynski equality~\cite{Kurchan2000,Tasaki2000,Campisi2009,Campisi2011}, the motion of the system is fully determined by a given path of a parameter.

  In the high-temperature limit $\beta = 0$, Eq.~\eqref{eq:can_IC} reduces to $\rho_{\Inter}\left( \lambda \right) = \id_\Sys\left( \lambda \right) / \mathcal{N}\left( \lambda \right) \otimes \sProj[\Con]{\lambda}$, where $\mathcal{N}\left( \lambda \right)$ is the dimensionality of the Hilbert space of $\Sys$ with fixed parameter $\lambda$.
  Thus, Eq.~\eqref{eq:gamma_2} reduces to $\gamma_{\lambda_\init \to \lambda_\fin} = \mathcal{N}\left( \lambda_\fin \right) / \mathcal{N}\left( \lambda_\init \right)$.
  In particular, when the dimensionality of the Hilbert space of $\Sys$ with fixed parameter $\lambda_\fin$ is equal to one with fixed parameter $\lambda_\init$, Eq.~\eqref{eq:gamma_2} reduces to unity whether $\calE$ is unital or not.

  We now argue for general values of $\beta$ that the quantity $\gamma_{\lambda_\init \to \lambda_\fin}$ is not necessarily unity for a unital map as $\gamma$ was for the cyclic process.
  When the time evolution $\calE$ is unital, namely the ``natural'' thermodynamic process defined in the previous subsection, the quantity $\gamma_{\lambda_\init \to \lambda_\fin}$ gives the ratio of the forward and the backward transition probabilities.
  When the time evolution $\calE$ is unital, completely positive and trace preserving, so is its adjoint $\calE^\dagger$.
  Therefore, we can regard the adjoint map $\calE^\dagger$ as another time evolution.
  Equation~\eqref{eq:qusi_trans_prob} indeed gives the backward transition probability that the state of $\Con$ changes from $\lambda_\fin$ to $\lambda_\init$:
  \begin{equation}
    q_{\lambda_\fin \to \lambda_\init}
    = \bTr{ \sProj[\Con]{\lambda_\init} \Emap[^\dagger]{\rho_{\Sys, \can}\left( \lambda_\fin \right) \otimes \sProj[\Con]{\lambda_\fin} } }. \label{eq:qusi_prob_unital}
  \end{equation}
  As can be seen from the calculation of a simple model in Sec.~\ref{sec:example}, the backward transition probability~\eqref{eq:qusi_prob_unital} is not necessarily equal to the forward one~\eqref{eq:trans_prob}.
  Therefore, the quantity $\gamma_{\lambda_\init \to \lambda_\fin}$ is not necessarily unity for a unital map.
  
  When the time evolution $\calE$ is not unital, incidentally, we cannot regard Eq.~\eqref{eq:qusi_trans_prob} as a transition probability; because the adjoint $\calE^\dagger$ of a non-unital map $\calE$ is not trace preserving, the sum of Eq.~\eqref{eq:qusi_trans_prob} over $\lambda_\init$ is not unity:
  \begin{align}
    \sum_{\lambda_\init} q_{\lambda_\fin \to \lambda_\init}
      &= \bTr{\rho_{\Sys, \can}\left( \lambda_\fin \right) \otimes \sProj[\Con]{\lambda_\fin} \Emap{\id_\Inter}} \\
      &= \bTr{\Emap[^\dagger]{\rho_{\Sys, \can}\left( \lambda_\fin \right) \otimes \sProj[\Con]{\lambda_\fin}}} 
      \neq 1,
  \end{align} 
  where $\id_\Inter := \sum_\lambda \id_\Sys\left( \lambda \right) \otimes \sProj[\Con]{\lambda}$ is the identity operator of $\Inter$.

  Finally, we show that Eq.~\eqref{eq:opened_Jar} reduces to the case of the cyclic process~\eqref{eq:Jarzyn_eq} when the control system $\Con$ has only one eigenstate.
  In this case, the state of $\Con$ cannot change from the initial state, and we thereby obtain $p_{\lambda_\init \to \lambda_\init} = 1$ and $\Delta F_\Inter\left( \lambda_\init, \lambda_\init \right) = 1$.
  Therefore, Eq.~\eqref{eq:open_P_after} reduces to
  \begin{multline}
    P_{\lambda_\init \to \lambda_\init}\left( W \right)
    = e^{-\beta W} \sum_j \delta\left( W - \Delta e_j \right) \\
    \times \bTr{\rho_{\Sys, \can}\left( \lambda_\init \right) \otimes \sProj[\Con]{\lambda_\init} \Emap[_j]{\id_\Sys\left( \lambda_\init \right) \otimes \sProj[\Con]{\lambda_\init}}} . 
  \end{multline}
  Since $\id_\Inter = \id_\Sys\left( \lambda_\init \right) \otimes \sProj[\Con]{\lambda_\init}$ and $\rho_{\Inter, \can} = \rho_{\Sys, \can}\left( \lambda_\init \right) \otimes \sProj[\Con]{\lambda_\init}$, this equation is equivalent to Eq.~\eqref{eq:cal_P(W)} in Sec.~\ref{sec:closed}.

\section{Coefficient $\gamma_{\lambda_\init \to \lambda_\fin}$ for a simple model \label{sec:example}}
  In this section, we evaluate the quantity $\gamma_{\lambda_\init \to \lambda_\fin}$ of Sec.~\ref{sec:opened} using a simple system.
  We suppose that the Hamiltonian of the simple system is given by
  \begin{gather}
    H_\Inter\left( \omega \right) = \sum_{\lambda = 0, 1} H_\Sys\left( \lambda; \omega \right) \otimes \sProj[\Con]{\lambda} , \\
    H_\Sys\left( \lambda; \omega \right) := \half{ \left( \lambda + 1 \right) \omega } \sigma_z^\Sys,
  \end{gather}
  where $\omega$ is level spacing and $\sigma_z^\Sys := \sProj[\Sys]{1} - \sProj[\Sys]{0}$.
  The canonical distribution $\rho_\Sys\left( \lambda \right)$ of $\Sys$ at the inverse temperature $\beta$ is given by
  \begin{equation}
    \rho_\Sys\left( \tilde{\beta} , \lambda \right)
      = \inv{1 + e^{\left( \lambda + 1 \right) \tilde{\beta}}} \sProj[\Sys]{1}
      + \inv{1 + e^{-\left( \lambda + 1 \right) \tilde{\beta}}} \sProj[\Sys]{0}
  \end{equation}
  where $\tilde{\beta} := \beta\omega$ is the dimensionless inverse temperature.

  \begin{figure}[tbp] 
    \includegraphics[width=\hsize]{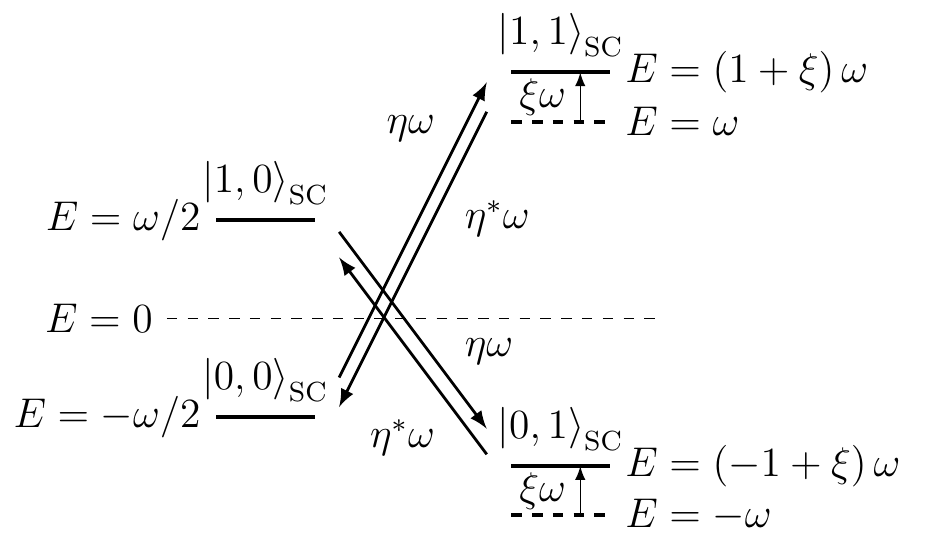}
    \caption{Illustration of the effective Hamiltonian of Eq.~\eqref{eq:eff_Ham}.
      The parameters $\xi\omega$ and $\eta\omega$ are real and complex numbers, respectively, and are represented to the potential energy and hopping amplitude, respectively. \label{fig:2x2_Ham}}
  \end{figure} 
  We consider the following measurement process $\calE_j$:
  \begin{gather}
    \Emap[_j]{\rho_\Inter} := M_j \rho_\Inter M_j^\dagger , \\
    M_j := \sum_{\lambda, \lambda^\prime} \sum_{\xbk{\lambda}, \ybk{\lambda^\prime}} \delta_{\Delta e_j, h_{\xbk{\lambda}} - h_{\ybk{\lambda^\prime}} } \Pi_{h_{\ybk{\lambda^\prime}}} U_\eff \Pi_{h_{\xbk{\lambda}}},
  \end{gather}
  where $\Pi_{h_\xbk{\lambda}}$ is a projection on the eigenvalue of $H_\Inter$.
  The effective time-evolution operator $U_\eff$ is given by tracing out the external system from the time evolution of the total system.
  It defines the effective Hamiltonian $H_\eff$ as in $U_\eff = e^{-i H_\eff T}$, where $T$ is the time duration of the measurement process $\calE_j$.
  The corresponding time evolution $\calE := \calE_j$ is unital, and therefore, is not a feedback process.
  For simplicity, let us suppose that the effective Hamiltonian $H_\eff$ is
  \begin{align}
    H_\eff\left( \omega, \xi, \eta \right) &:= H_\Inter\left( \omega \right) + V_\eff\left( \xi, \omega \right) + H_\hop\left( \eta, \omega \right) , \label{eq:eff_Ham} \\
    V_\eff\left( \xi, \omega \right) &:= \xi\omega \id_\Sys \otimes \sProj[\Con]{1}, \\
    H_\hop\left( \eta, \omega \right) &:= \sigma_x^\Sys \otimes \omega\left( \eta \sKetbra[\Con]{1}{0} + \eta^\ast \sKetbra[\Con]{0}{1} \right),
  \end{align}
  where $V_\eff\left( \xi, \omega \right)$ and $H_\hop\left( \eta, \omega \right)$ are potential and hopping terms, the parameters $\xi$ and $\eta$ are real and complex numbers, respectively (Fig.~\ref{fig:2x2_Ham}), and $\sigma_x^\Sys := \sKetbra[\Sys]{1}{0} + \sKetbra[\Sys]{0}{1}$.
  In other words, the unitary operator $U_\eff$ is given by
  \begin{equation}
    U_\eff \left( \xi, \eta, \tilde{T} \right)
    := \exp\left( -i \tilde{T} H_\eff\left( 1, \xi, \eta \right) \right) ,
    \label{eq:eff_U}
  \end{equation}
  where $\tilde{T} := \omega T / \hbar$ is the dimensionless time duration.

  Then, Eqs.~\eqref{eq:trans_prob} and \eqref{eq:qusi_trans_prob} are given by
  \begin{gather}
    p_{\lambda_\init \to \lambda_\fin}\left( \xi, \eta, \tilde{\beta} , \tilde{T} \right)
      \!=\! \frac{p_{\lambda_\init \to \lambda_\fin}^{(1)}\left( \xi, \eta, \tilde{T} \right)}{1 + e^{\left( \lambda_\init + 1 \right) \tilde{\beta}}}
      +  \frac{p_{\lambda_\init \to \lambda_\fin}^{(0)}\left( \xi, \eta, \tilde{T} \right)}{1 + e^{-\left( \lambda_\init + 1 \right) \tilde{\beta}}} , \label{eq:p_eff} \\
    q_{\lambda_\fin \to \lambda_\init}\left( \xi, \eta, \tilde{\beta}, \tilde{T} \right)
      \!=\! \frac{q_{\lambda_\fin \to \lambda_\init}^{(1)}\left( \xi, \eta, \tilde{T} \right)}{1 + e^{\left( \lambda_\fin + 1 \right) \tilde{\beta}}}
      + \frac{q_{\lambda_\fin \to \lambda_\init}^{(0)}\left( \xi, \eta, \tilde{T} \right)}{1 + e^{-\left( \lambda_\fin + 1 \right) \tilde{\beta}}} , \label{eq:q_eff}
  \end{gather}
  with
  \begin{widetext}
    \begin{align}
      p_{\lambda_\init \to \lambda_\fin}^{(n)}\left( \xi, \eta, \tilde{T} \right)
        &:= \bTr{\sProj[\Con]{\lambda_\fin} U_\eff\left( \xi, \eta, \tilde{T} \right) \left( \sProj[\Sys]{n} \otimes \sProj[\Con]{\lambda_\init} \right) U_\eff^\dagger\left( \xi, \eta, \tilde{T} \right) }, \label{eq:pn}\\
      q_{\lambda_\fin \to \lambda_\init}^{(n)}\left( \xi, \eta, \tilde{T} \right)
        &:= \bTr{\sProj[\Con]{\lambda_\init} U_\eff^\dagger\left( \xi, \eta, \tilde{T} \right) \left( \sProj[\Sys]{n} \otimes \sProj[\Con]{\lambda_\fin} \right) U_\eff\left( \xi, \eta, \tilde{T} \right) } \label{eq:qn}
    \end{align}
  \end{widetext}
  for $n = 0, 1$.

  We first show $\gamma_{0 \to 0} = \gamma_{1 \to 1} = 1$ and $\gamma_{1 \to 0} = 1 / \gamma_{0 \to 1}$.
  The projective operators $\sProj[\Sys]{n}$ $(n = 0, 1)$ and $\sProj[\Con]{\lambda}$ $(\lambda = 0, 1)$ are invariant with respect to a unitary operator $S_1 := \id_\Sys \otimes \left( \sProj[\Con]{1} + e^{-i\chi} \sProj[\Con]{0} \right)$ for any real number $\chi$ and the time reversal operator $\Theta$, which is anti-unitary operator, and the unitary operator~\eqref{eq:eff_U} satisfies
  \begin{align}
    S_1^\dagger U_\eff\left( \xi, \eta, \tilde{T} \right) S_1 &= U_\eff\left( \xi, e^{i\chi}\eta, \tilde{T} \right) , \\
    \Theta^\dagger U_\eff\left( \xi, \eta, \tilde{T} \right) \Theta &= U_\eff^\dagger\left( \xi, \eta^\ast, \tilde{T} \right) .
  \end{align}
  Applying $S_1$ and $\Theta$ in Eqs.~\eqref{eq:pn} and \eqref{eq:qn}, we obtain
  \begin{align}
    p_{\lambda_\init \to \lambda_\fin}^{(n)}\left( \xi, \eta, \tilde{T} \right)
      &= p_{\lambda_\init \to \lambda_\fin}^{(n)}\left( \xi, e^{i\chi}\eta, \tilde{T} \right), \label{eq:p_mb} \\
    p_{\lambda_\init \to \lambda_\fin}^{(n)}\left( \xi, \eta, \tilde{T} \right)
      &= q_{\lambda_\init \to \lambda_\fin}^{(n)}\left( \xi, \eta^\ast, \tilde{T} \right) . \label{eq:q_pb}
  \end{align}
  For a unitary operator $S_2 := \sigma_x^\Sys \otimes \id_\Con$, we also have
  \begin{align}
    S_2^\dagger U_\eff\left( \xi, \eta, \tilde{T} \right) S_2 &=  U_\eff^\dagger\left( -\xi, -\eta, \tilde{T} \right) , \\
    S_2^\dagger \left( \sProj[\Sys]{n} \otimes \sProj[\Con]{\lambda} \right) S_2 &= \sProj[\Sys]{1 - n} \otimes \sProj[\Con]{\lambda}  
  \end{align}
  for $n, \lambda = 0, 1$, and hence
  \begin{equation}
    p_{\lambda_\init \to \lambda_\fin}^{(n)}\left( \xi, \eta, \tilde{T} \right)
      = q_{\lambda_\init \to \lambda_\fin}^{(1 - n)}\left( -\xi, -\eta, \tilde{T} \right) .
    \label{eq:p_ma_mb}
  \end{equation}
  Combining Eqs.~\eqref{eq:p_mb}, \eqref{eq:q_pb} and \eqref{eq:p_ma_mb}, we thus obtain
  \begin{align}
    p_{\lambda_\init \to \lambda_\fin}^{(n)}\left( \xi, \Abs{\eta}, \tilde{T} \right)
      &= p_{\lambda_\init \to \lambda_\fin}^{(n)}\left( \xi, \eta, \tilde{T} \right) \\
      &= q_{\lambda_\init \to \lambda_\fin}^{(n)}\left( \xi, \eta, \tilde{T} \right) \\
      &= p_{\lambda_\init \to \lambda_\fin}^{(1 - n)}\left( -\xi, \eta, \tilde{T} \right) .
  \end{align}
  Therefore, we obtain
  \begin{multline}
    p_{\lambda_\init \to \lambda_\fin}\left( \xi, \eta, \tilde{\beta} , \tilde{T} \right)
      = q_{\lambda_\init \to \lambda_\fin}\left( \xi, \eta, \tilde{\beta} , \tilde{T} \right) \\
      = \frac{p_{\lambda_\init \to \lambda_\fin}^{(0)}\left( -\xi, \Abs{\eta}, \tilde{T} \right)}{1 + e^{\left( \lambda_\init + 1 \right) \tilde{\beta}}}
      +  \frac{p_{\lambda_\init \to \lambda_\fin}^{(0)}\left( \xi, \Abs{\eta}, \tilde{T} \right)}{1 + e^{-\left( \lambda_\init + 1 \right) \tilde{\beta}}} ,
    \label{eq:p_eff_2} 
  \end{multline}
  and hence
  \begin{align}
    \gamma_{\lambda_\init \to \lambda_\fin}\left( \xi, \eta, \tilde{\beta} , \tilde{T} \right)
      &= \frac{q_{\lambda_\fin \to \lambda_\init}\left( \xi, \eta, \tilde{\beta}, \tilde{T} \right)}{p_{\lambda_\init \to \lambda_\fin}\left( \xi, \eta, \tilde{\beta} , \tilde{T} \right)}  \\
      &= \frac{p_{\lambda_\fin \to \lambda_\init}\left( \xi, \eta, \tilde{\beta}, \tilde{T} \right)}{p_{\lambda_\init \to \lambda_\fin}\left( \xi, \eta, \tilde{\beta} , \tilde{T} \right)}  .
    \label{eq:g_eff}
  \end{align}
  As can be seen from Eq.~\eqref{eq:g_eff}, the quantity $\gamma_{\lambda_\init \to \lambda_\fin}$ is always equal to unity if $\lambda_\init = \lambda_\fin$, that is, $\gamma_{0 \to 0} = \gamma_{1 \to 1} = 1$, and $\gamma_{1 \to 0} = 1 / \gamma_{0 \to 1}$.

\begin{figure*}[tbp]
    \begin{tabular}{cc}
      \multicolumn{2}{c}{\includegraphics[width=8cm]{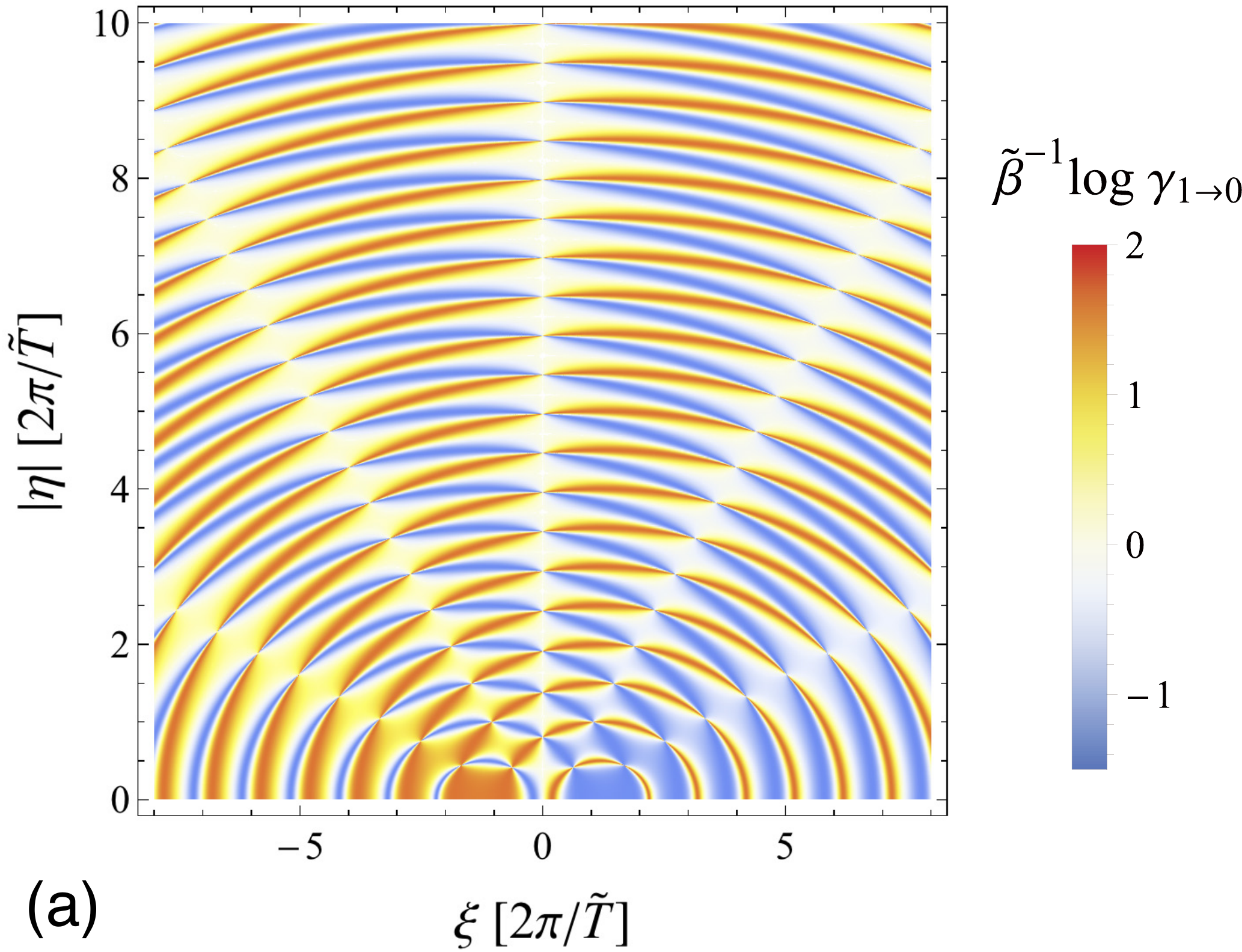}} \\
      \includegraphics[width=8cm]{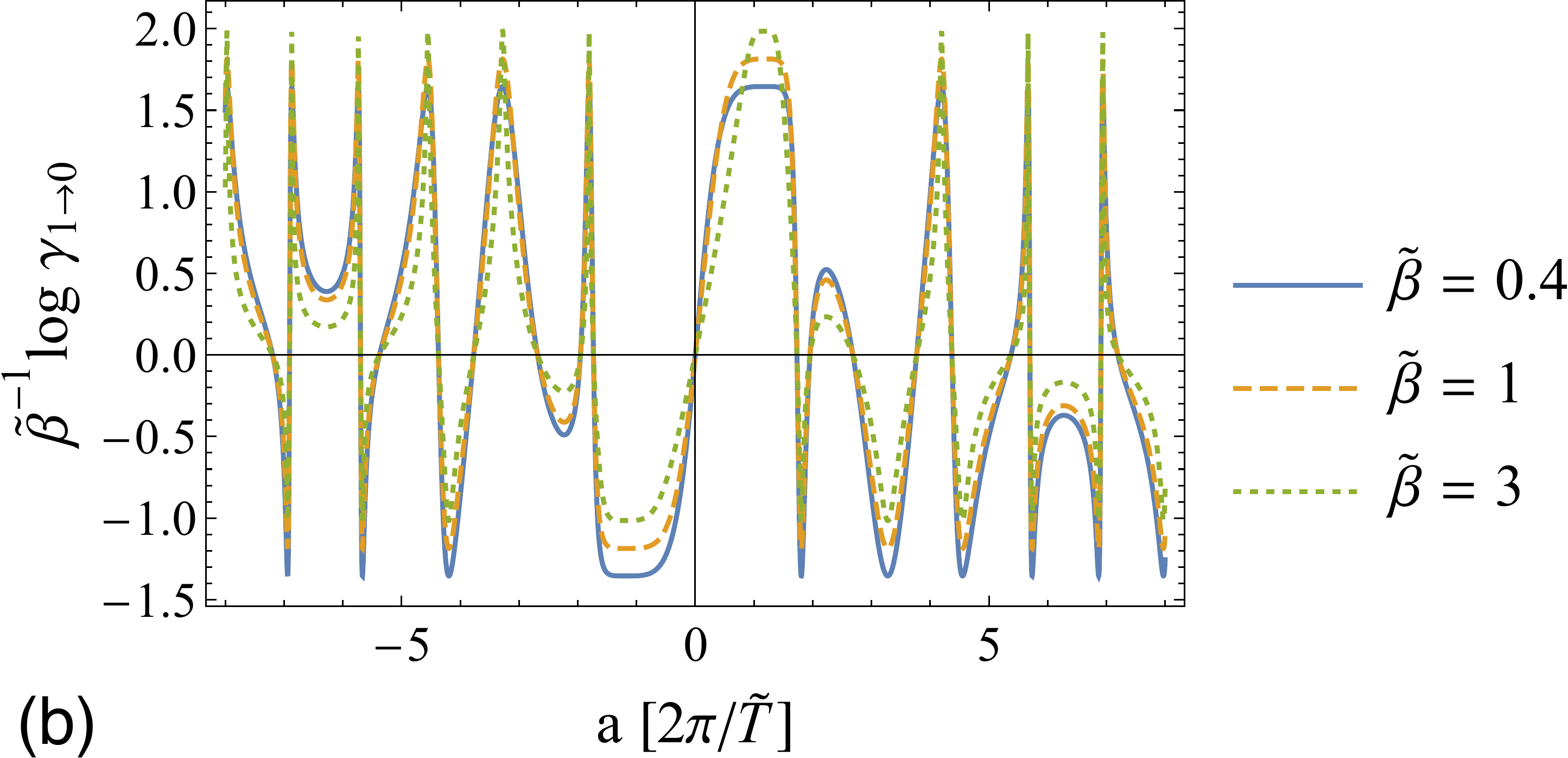} &
      \includegraphics[width=8cm]{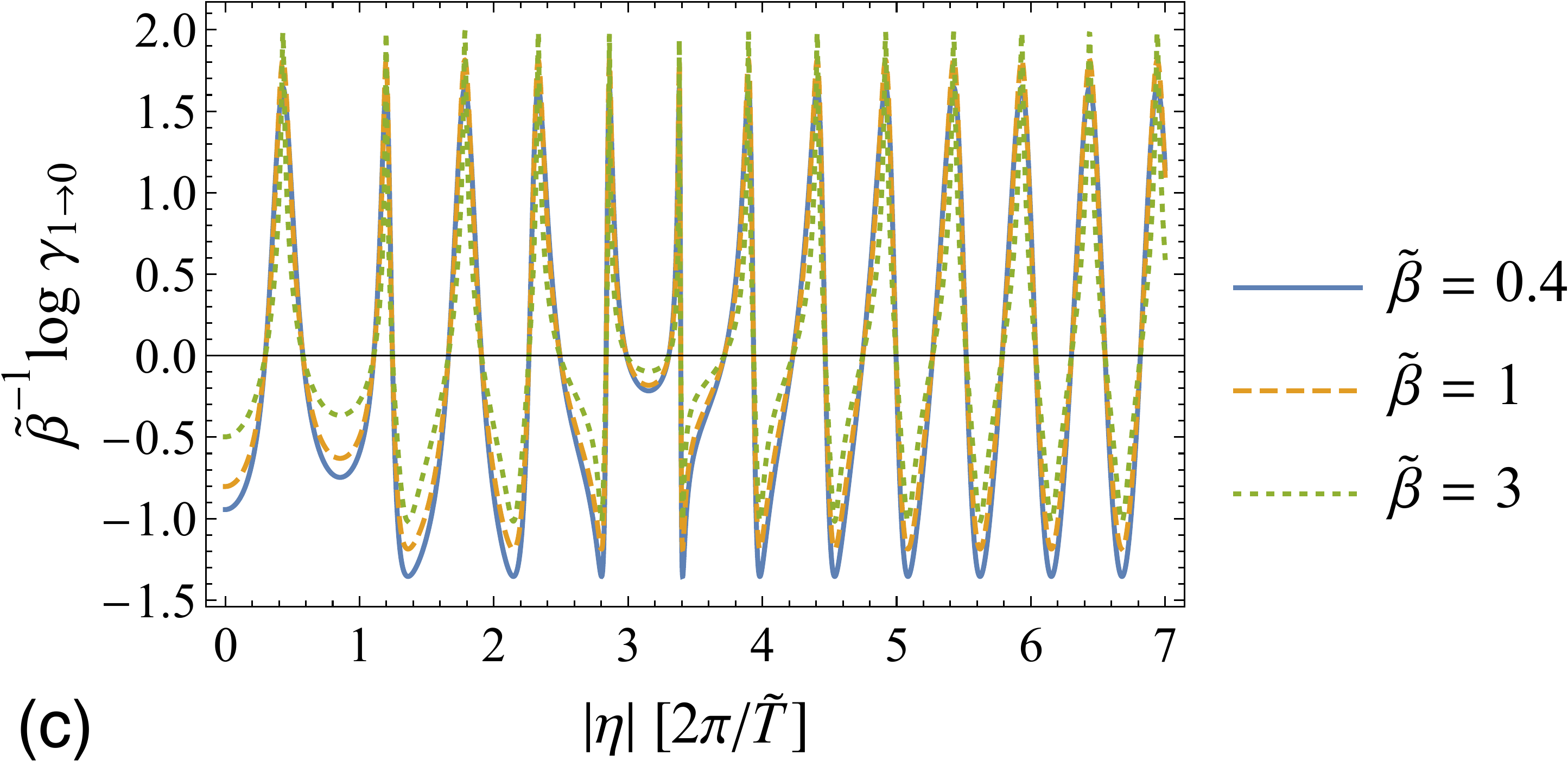} \\
    \end{tabular}
    \caption{(Color online) (a) Dependence of $\tilde{\beta}^{-1}\log\gamma_{1 \to 0}$ on the parameters $\xi$ and $\Abs{\eta}$ for $\tilde{T} = 5 $ and $\tilde{\beta} = 0.4$. Its cross sections for (b) $\Abs{\eta} = 4\pi/5$ and (c) $\xi = 6\pi/5$. The unity of $\gamma_{1 \to 0}$ (namely the zero of $\log\gamma_{1 \to 0}$) does not depend on the inverse temperature $\tilde{\beta}$. \label{fig:gamma}}
  \end{figure*}

  Let us now find $\gamma_{1 \to 0}$.
  Calculating $p_{1 \to 0}^{(0)}\left( \xi, \Abs{\eta}, \tilde{T} \right)$ and $p_{0 \to 1}^{(0)}\left( \xi, \Abs{\eta}, \tilde{T} \right)$, we obtain
  \begin{align}
    p_{1 \to 0}^{(0)}\left( \xi, \Abs{\eta}, \tilde{T} \right)
      &= p_{0 \to 1}^{(0)}\left( -\xi, \Abs{\eta}, \tilde{T} \right) \\
      &= \Abs{\eta}^2 \tilde{T}^2 \sinc^2\left( f\left( \xi, \Abs{\eta} \right) \tilde{T} \right) ,
  \end{align}
  with 
  \begin{gather}
    \sinc\left( x \right) := \frac{\sin\left( x \right)}{x}, \\ 
    f\left( \xi, \Abs{\eta} \right) := \sqrt{\Abs{\eta}^2 + \left( \frac{3 - 2\xi}{4} \right)^2}
  \end{gather}
  and hence
  \begin{widetext}
    \begin{align}
      p_{1 \to 0}\left( \xi, \Abs{\eta}, \tilde{\beta} , \tilde{T} \right)
        &= \frac{\Abs{\eta}^2 \tilde{T}^2 \sinc^2\left( f\left( -\xi, \Abs{\eta} \right) \tilde{T} \right)}{1 + e^{2\tilde{\beta}}}
        + \frac{\Abs{\eta}^2 \tilde{T}^2 \sinc^2\left( f\left( \xi, \Abs{\eta} \right) \tilde{T} \right)}{1 + e^{-2\tilde{\beta}}} , \\
      p_{0 \to 1}\left( \xi, \Abs{\eta}, \tilde{\beta} , \tilde{T} \right)
        &= \frac{\Abs{\eta}^2 \tilde{T}^2 \sinc^2\left( f\left( \xi, \Abs{\eta} \right) \tilde{T} \right)}{1 + e^{\tilde{\beta}}}
        + \frac{\Abs{\eta}^2 \tilde{T}^2 \sinc^2\left( f\left( -\xi, \Abs{\eta} \right) \tilde{T} \right)}{1 + e^{-\tilde{\beta}}} .
    \end{align}
  \end{widetext}
  We thereby plotted $\gamma_{1 \to 0}$ in Fig.~\ref{fig:gamma}.
  The quantity $\gamma_{1 \to 0}$ fluctuates around unity wildly, even exceeding unity often.
  This fluctuation of $\gamma_{\lambda_\init \to \lambda_\fin}$ should be detectable in experiments and provide evidence for the present approach.

  For $\xi = 0$, we find $p_{1 \to 0}\left( 0, \eta, \tilde{\beta} , \tilde{T} \right) = p_{0 \to 1}\left( 0, \eta, \tilde{\beta} , \tilde{T} \right)$ and $\gamma_{1 \to 0}\left( 0, \eta, \tilde{\beta}, \tilde{T} \right) = 1$.
  The other conditions for $\gamma_{1 \to 0} = 1$ is $\tilde{\beta} = 0$ or
  \begin{equation}
    \sinc^2\left( f\left( \xi, \Abs{\eta} \right) \tilde{T} \right)
    = \sinc^2\left( f\left( -\xi, \Abs{\eta} \right) \tilde{T} \right),
  \end{equation}
  which does not depend on the inverse temperature $\tilde{\beta}$~(Fig.~\ref{fig:gamma}b, c).

\section{Conclusion\label{sec:Conclustion}}
  In the present paper, we have shown that the variance of the energy transferred to the external system diverges when the dynamics of the internal quantum system is approximated to a unitary.
  Because of this, the work extraction under the assumption of a unitary dynamics of the internal system is unsuitable for the thermodynamics of a microscopic quantum system.
  We claim that the work extraction from the internal quantum system should be described as a quantum measurement process introduced by Hayashi and Tajima~\cite{Hayashi2015,TajimaThesis} and have applied this formulation to the quantum Jarzynski equality.

  In the present paper, we have assumed that the measurement process can measure the energy loss of the internal system without errors; in other words, it can convert all of the energy loss into the work.
  However, since the actual external system is a thermodynamic system, the energy transfer may be separated into the work and the heat.
  The measurement process corresponding to this should be an incomplete process in which the extracted work is not equal to the energy loss; the heat generated into the external system should be regarded as the difference of the energy loss and the extracted work.
  It will be an important future work to consider the incomplete measurement process in order to understand the work and the heat in quantum thermodynamics.

\appendix
\section{Calculation of Eqs.~\eqref{eq:avr_h} and \eqref{eq:avr_h_2} \label{app:app_B}}
  We here show details of the calculation of Eqs.~\eqref{eq:expect_h} and \eqref{eq:expect_h2}.
  To calculate them, we introduce to the matrix representation of Eqs.~\eqref{eq:arbitrary_density} and \eqref{eq:approx_K}: 
  \begin{gather}
    \rho_\Inter := \begin{pmatrix} p && q \\ q^\ast && 1 - p \end{pmatrix} ,
    \label{eq:matrix_arbitrary_density} \\
    K_{n, n^\prime}
    = \begin{pmatrix} 
        \delta_{n, n^\prime} u_{0,0} && \delta_{n - 1, n^\prime} u_{0,1} \\
        \delta_{n + 1, n^\prime} u_{1,0} && \delta_{n, n^\prime} u_{1,1}
    \end{pmatrix} .
  \end{gather}
  We find
    \begin{equation}
      \bar{K}_j := \sum_{m = 1}^M K_{j, m}
        = \begin{cases}
          \begin{pmatrix} 0 & 0 \\ u_{1,0} & 0 \end{pmatrix}
            &\text{for $j = 0$} \\
          \begin{pmatrix} u_{0,0} & 0 \\ u_{1,0} & u_{1,1} \end{pmatrix}
            &\text{for $j = 1$} \\
          \begin{pmatrix} u_{0,0} & u_{0,1} \\ u_{1,0} & u_{1,1} \end{pmatrix}
            &\text{for $2 \leq j \leq M - 1$} \\
          \begin{pmatrix} u_{0,0} & u_{0,1} \\ 0 & u_{1,1} \end{pmatrix}
            &\text{for $j = M$} \\
          \begin{pmatrix} 0 & u_{0,1} \\ 0 & 0 \end{pmatrix}
            &\text{for $j = M + 1$} \\
          0 & \text{otherwise}
        \end{cases}
    \end{equation}
    and thereby obtain
  \begin{widetext}
    \begin{equation}
      \bar{K}_j^\dagger \bar{K}_j
        = \begin{cases}
          \begin{pmatrix} \Abs{u_{1,0}}^2 & 0 \\ 0 & 0 \end{pmatrix}
            &\text{for $j = 0$} \\
          \begin{pmatrix} \Abs{u_{0,0}}^2 + \Abs{u_{1,0}}^2 & u_{1,0}^\ast u_{1,1} \\
            u_{1,0} u_{1,1}^\ast & \Abs{u_{1,1}}^2 \end{pmatrix}
            &\text{for $j = 1$} \\
          \begin{pmatrix} \Abs{u_{0,0}}^2 + \Abs{u_{1,0}}^2 & u_{0,0}^\ast u_{0,1} + u_{1,0}^\ast u_{1,1} \\
            u_{0,0} u_{0,1}^\ast + u_{1,0} u_{1,1}^\ast & \Abs{u_{0,1}}^2 + \Abs{u_{1,1}}^2 \end{pmatrix}
            &\text{for $2 \leq j \leq M - 1$} \\
          \begin{pmatrix} \Abs{u_{0,0}}^2 & u_{0,0}^\ast u_{0,1} \\
            u_{0,0} u_{0,1}^\ast & \Abs{u_{0,1}}^2 + \Abs{u_{1,1}}^2 \end{pmatrix}
            &\text{for $j = M$} \\
          \begin{pmatrix} 0 & 0 \\ 0 & \Abs{u_{0,1}}^2 \end{pmatrix}
            &\text{for $j = M + 1$} \\
          0 & \text{otherwise}
        \end{cases}
    \end{equation}
    Using Eqs.~\eqref{eq:a^2}, \eqref{eq:d^2} and \eqref{eq:ad}, we obtain
    \begin{gather}
      \sum_j j \bar{K}_j^\dagger \bar{K}_j
        = \half{1} M\left( M + 1 \right)
        + \begin{pmatrix}
          -\Abs{u_{1,0}}^2 M & u_{0,0}^\ast u_{0,1} \left( M - 1 \right) \\
          u_{0,0} u_{0,1}^\ast \left( M - 1 \right) & \Abs{u_{0,1}}^2 M
          \end{pmatrix} , \label{eq:jKK} \\
      \sum_j j^2 \bar{K}_j^\dagger \bar{K}_j
      = \frac{1}{6} M\left( M + 1 \right)\left( 2M + 1 \right)
        + \begin{pmatrix}
          -\Abs{u_{1,0}}^2 M^2 & u_{0,0}^\ast u_{0,1} \left( M^2 - 1 \right) \\
          u_{0,0} u_{0,1}^\ast \left( M^2 - 1 \right) & \Abs{u_{0,1}}^2 \left( M^2 + 2M \right)
          \end{pmatrix} \label{eq:j2KK} .
    \end{gather}
    Inserting Eqs.~\eqref{eq:matrix_arbitrary_density}, \eqref{eq:jKK} and \eqref{eq:j2KK} into Eqs.~\eqref{eq:expect_h} and \eqref{eq:expect_h2}, we obtain
    \begin{gather}
      \Avr{h}
      = \Delta \left[ \half{1} \left( M + 1 \right) + \Abs{u_{0,1}}^2 \left( 1 - p \right) - \Abs{u_{1,0}}^2 p +2\left( 1 - M^{-1} \right) \text{Re}\left( u_{0,0} u_{0,1}^\ast q \right) \right] ,
      \label{eq:avr_h_app} \\
      \Avr{h^2} 
        = \Delta^2 \left[ \frac{1}{6} \left( M + 1 \right)\left( 2M + 1 \right) + \Abs{u_{0,1}}^2 \left( M + 2 \right) \left( 1 - p \right) - \Abs{u_{1,0}}^2 M p + 2\left( M - M^{-1} \right) \text{Re}\left( u_{0,0} u_{0,1}^\ast q \right) \right] .
        \label{eq:avr_h_2_app}
    \end{gather}
  
    Now, we arrange Eqs.~\eqref{eq:avr_h_app} and \eqref{eq:avr_h_2_app} using the average of the energy transfer $\Avr{w}$.
    The expectation of energy with respect to the initial state~\eqref{eq:init_psi} is given by
    \begin{equation}
      \bTr{H_\Ext \rho_\Ext} = \half{\Delta} \left( M + 1 \right) .
      \label{eq:init_psi_energy}
    \end{equation}
    Combining Eqs.~\eqref{eq:two_lev_wj}, \eqref{eq:avr_h_app} and \eqref{eq:init_psi_energy}, we obtain the expectation of $w$ in the form
    \begin{align}
      \Avr{w} 
      &= \Avr{h} - \bTr{H_\Ext \rho_\Ext} \\
      &= \Delta \left[ \Abs{u_{0,1}}^2 \left( 1 - p \right) - \Abs{u_{1,0}}^2 p +2\left( 1 - M^{-1} \right) \text{Re}\left( u_{0,0} u_{0,1}^\ast q \right) \right] .
    \end{align}
  \end{widetext}
  Therefore, Eqs.~\eqref{eq:avr_h_app} and \eqref{eq:avr_h_2_app} reduce to Eqs.~\eqref{eq:avr_h} and \eqref{eq:avr_h_2}.

\section{Calculation of Eq.~\eqref{eq:d_R-R} \label{app:app_A}}
  Let us find Eq.~\eqref{eq:d_R-R}.
  The eigenvalues of $\Kmap{\rho_\Inter}$ and $\rho_\Inter$ are given by
  \begin{equation}
    \lambda_\pm := \half{1} \pm R , \quad
    \lambda^\prime_\pm := \half{1} \pm R^\prime , 
  \end{equation}
  where $R$ and $R^\prime$ are defined in Eqs.~\eqref{eq:R} and \eqref{eq:R^prime}, respectively.
  The diagonal matrices $\Lambda$ and $\Lambda^\prime$ are given by
  \begin{equation}
    \Lambda := \begin{pmatrix} \lambda_{+} && 0 \\ 0 && \lambda_{-} \end{pmatrix} , \quad
    \Lambda^\prime := \begin{pmatrix} \lambda^\prime_{+} && 0 \\ 0 && \lambda^\prime_{-} \end{pmatrix} .
    \label{eq:diag} 
  \end{equation}

  We parameterize the arbitrary unitary matrix $\tilde{U}_\Inter$ as
  \begin{equation} 
    \tilde{U}_\Inter := e^{i\phi} \begin{pmatrix}
        e^{i\psi_1} \cos\theta && e^{i\psi_2} \sin\theta \\
        -e^{-i\psi_2} \sin\theta && e^{-i\psi_1} \cos\theta
      \end{pmatrix} ,
    \label{eq:2_unitary}
  \end{equation}
  where $\phi$, $\theta$, $\psi_1$ and $\psi_2$ are real parameters.
  Combining Eqs.~\eqref{eq:diag} and \eqref{eq:2_unitary}, we obtain
  \begin{widetext}
    \begin{equation}
      \tilde{U}_\Inter \Lambda^\prime \tilde{U}_\Inter^\dagger - \Lambda \\
      = \begin{pmatrix}
        R^\prime \cos 2\theta - R && -e^{i\left( \psi_1 + \psi_2 \right)} R^\prime \sin 2\theta \\
        -e^{-i\left( \psi_1 + \psi_2 \right)} R^\prime \sin 2\theta && -\left( R^\prime \cos 2\theta - R \right) 
        \end{pmatrix} ,
    \end{equation}
    where eigenvalues are given by $\pm\sqrt{{R^\prime}^2 + R^2 - 2 R R^\prime \cos 2\theta}$.
    We thereby obtain
    \begin{equation}
      \half{1} \Tr \Abs{ \tilde{U}_\Inter \Lambda^\prime \tilde{U}_\Inter^\dagger - \Lambda} 
      = \sqrt{{R^\prime}^2 + R^2 - 2 R R^\prime \cos 2\theta} .
    \end{equation}
    Therefore, we obtain Eq.~\eqref{eq:d_R-R}.
  \end{widetext}


%

\end{document}